\documentclass[11pt,titlepage]{article}

\usepackage{setspace} 

\usepackage{amsmath}

\usepackage{graphicx}

\usepackage[round,numbers,sort&compress]{natbib} 
\title{Ionic partition and transport in multi-ionic channels: A Molecular Dynamics Simulation study of the OmpF bacterial porin}

\author{Jordi Faraudo and Carles Calero \\
        Institut de Ci\`encia de Materials de Barcelona (ICMAB-CSIC)\\
        Campus de la UAB, E-08193 Bellaterra, Spain\\
        \and Marcel Aguilella-Arzo\thanks{All authors have contributed equally to this work. Address correspondence to M. Aguilella-Arzo (arzo@fca.uji.es) }\\
        Biophysics Group, Department of Physics,\\
        Universitat Jaume I, 12080 Castell\'o, Spain.\\}

\date{}

\begin{document}
\maketitle

\abstract{We performed all-atom molecular dynamics simulations studying the partition of ions and the ionic current through the bacterial porin OmpF and two selected mutants. The study is motivated by new interesting experimental findings concerning their selectivity and conductance behaviour at neutral pH. The mutations considered here are designed to study the effect of removal of negative charges present in the constriction zone of the wild type OmpF channel (which contains on one side a cluster with three positive residues and on the other side two negatively charged residues). Our results show that these mutations induce an exclusion of cations from the constriction zone of the channel, substantially reducing the flow of cations. In fact, the partition of ions inside the mutant channels is strongly inhomogeneous, with regions containing excess of cations and regions containing excess of anions. Interestingly, the overall number of cations inside the channel is larger than the number of anions in the two mutants, as in the OmpF wild type channel. We found that the differences in ionic charge inside these channels are justified by the differences in electric charge between the wild type OmpF and the mutants, following an electroneutral balance.}

\clearpage

\section*{Introduction}

The continuous increase in computer power has made possible the use of Molecular Dynamics (MD) Simulations as a kind of computational microscope to obtain dynamic images of biomolecular systems and processes with atomistic resolution. Relevant recent examples include the interaction of a nascent protein with the ribosome \cite{Trabuco2010} or electrophoresis of DNA \cite{Aksimentiev2010}. It is also possible to obtain useful insights in problems which were thought to be computationally intractable, such as protein folding \cite{Freddolino2010}.

The permeation of ions across protein channels is one of the biophysical problems to which MD simulations are becoming increasingly relevant \cite{ReviewSoftMatter}. Equilibrium MD simulations including free energy calculations and analysis of ion binding sites have revealed many interesting insights on channel permeation \cite{Guidoni1999,Phale2001,Danelon2003,Noskov2004}. Although most of these early MD simulations are based on classical force fields appropiately parametrized, some MD simulations of ions inside protein channels are based on \emph{ab initio} quantum mechanical calculations for the ion-protein interaction \cite{Guidoni2002}.

The recent development of new algorithms \cite{NAMD} for large-scale MD simulations has made possible non-equilibrium simulations for direct conductance calculations \cite{Aksimentiev}, revealing important insights on the ion conductance on a molecular level. These calculations still remain computationally very expensive, and only a few protein channels have been simulated under applied electrical fields. Relevant examples are the $\alpha-$hemolysin \cite{Aksimentiev} toxin from \emph{Staphylococcus aureus} or the two major porins of \emph{Escherichia coli}, OmpF \cite{Chimerel2008,Ulrich2009} and OmpC \cite{Biro2010}. It has to be emphasized that wide channels such as those toxins and bacterial porins are designed by nature for the transport of metabolites and other molecules much larger than small inorganic ions. However, ionic conductivity in these channels is also of great interest as model systems to test our understanding of the physical principles of ionic transport in channels and nanoscale electrostatics \cite{Review}. In fact, many bacterial porins can be modified genetically, offering broader perspectives to test theories and models on ion permeation.

\begin{figure}[htp]
   \begin{center}   
       \includegraphics*[width=12cm]{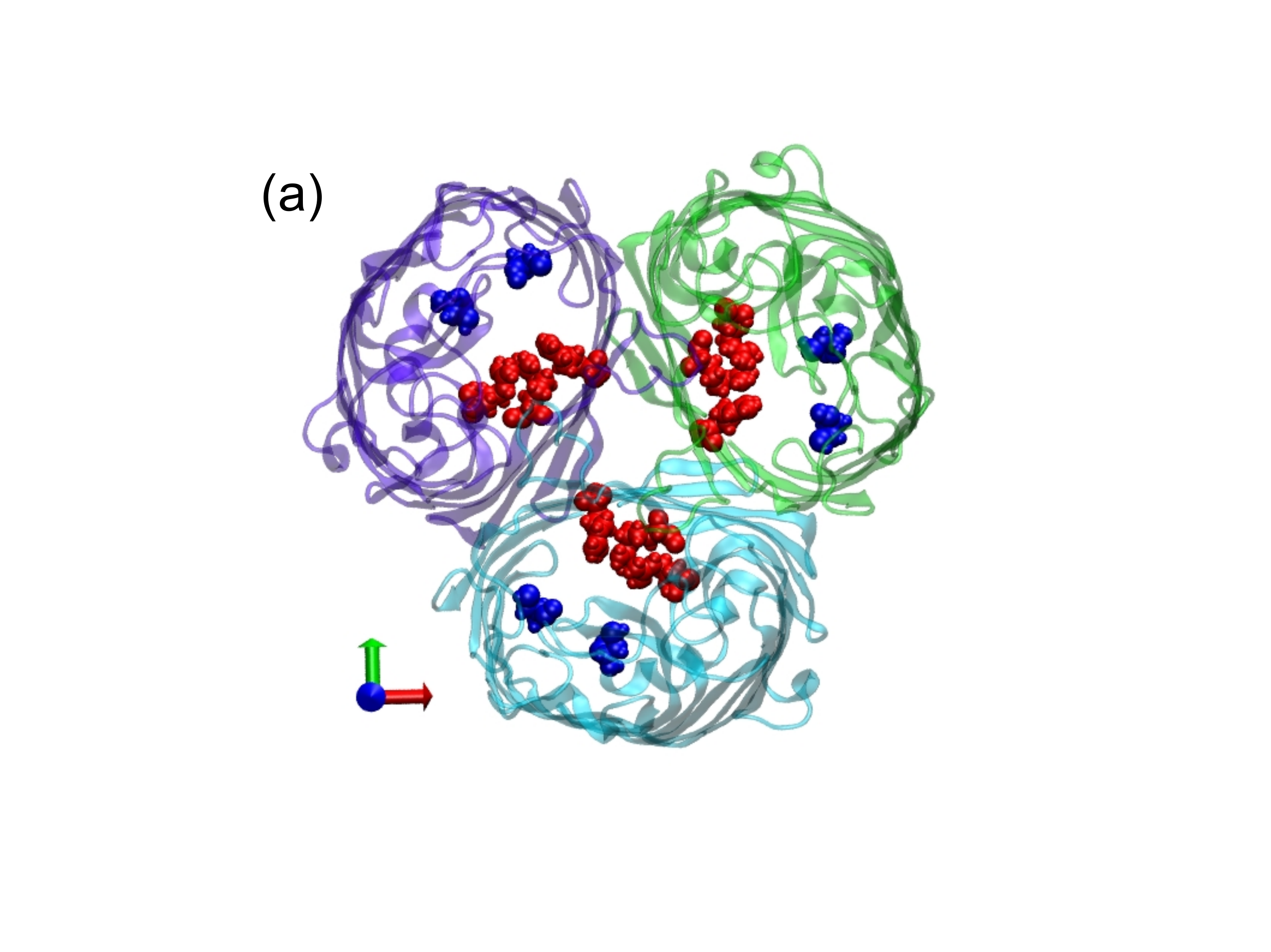}
	\includegraphics*[width=13cm]{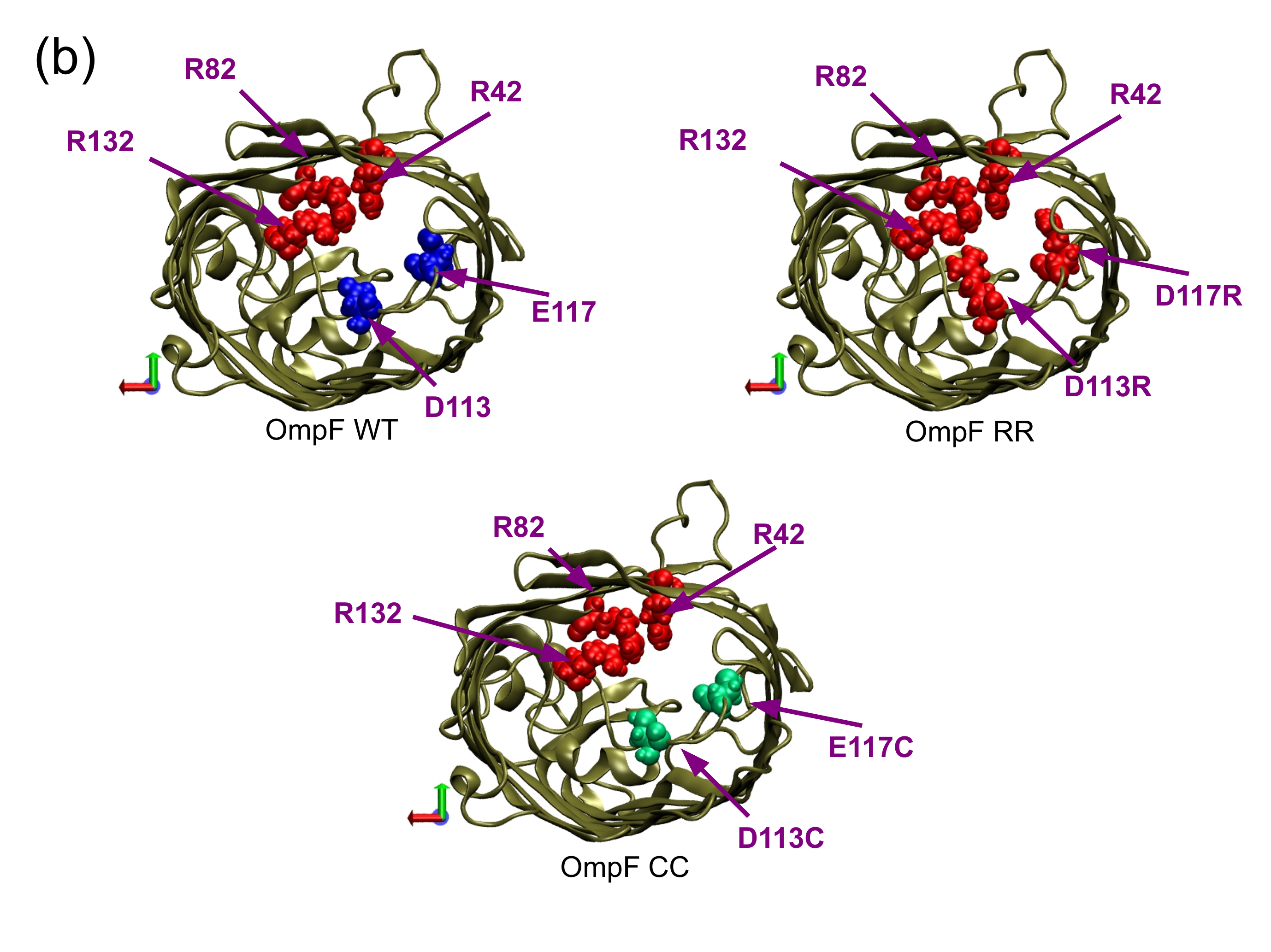}
       \caption{\label{channels} (a) Top view of the OmpF trimer, highlighting the residues located in the constriction zone. (b) Top view of one monomer for the wild-type (OmpF-WT) and the two mutants (CC and RR) considered in this work. Residues indicated in blue are negativelly charged, residues in red are positively charged and residues in green do not have net charge at neutral pH.}
    \end{center}
 \end{figure}

In this work, we present a large-scale non-equilibrium MD simulation study of the OmpF channel and two of its mutants, motivated by recent experimental results \cite{Alcaraz2009,Alcaraz2010}. The structure of the OmpF channel and the two mutants considered here (which are the same considered in Ref.\cite{Alcaraz2010}) is illustrated in Figure \ref{channels}. The crystallographical structure of the wild type OmpF channel is well known from long ago \cite{Cowan,PDB}. It is made of three identical monomerical nanopores, each one assembling into a large 16-stranded antiparallel $\beta$-barrel structure enclosing the transmembrane pore (Figure \ref{channels}a). Each aqueous pore has a diameter between 1-4 nm, being constricted around half-way through the membrane by a long loop. At neutral pH and in presence of monovalent electrolyte it has a slight cationic selectivity which has been analyzed in detail with many different theoretical approaches including mean field continuum models based on Poisson-Nernst-Planck, Brownian Dynamics Simulations and Molecular Dynamics (MD) simulations \cite{Phale2001,Danelon2003,Aguilella2007,Roux,Roux2002b}. A remarkable feature of this channel is the presence of a substantial transversal electrical field at the constriction zone, generated by clusters of positively and negatively charged residues (see Figure \ref{channels}a). This transversal field may serve as an aid to the permeation of dipolar solutes, which is the natural function of the OmpF porin \cite{Cowan1992,Cowan1994}. Simulation \cite{Roux,Roux2002b} and X-ray studies \cite{Roux2010} demonstrate that this transversal field affects the translocation of ions across this multi-ionic channel by generating separate pathways for cations and anions in a screw-like fashion across the $\beta$-barrel. This pathway separation is most clear at the constriction zone, where the cationic pathway is close to the negatively charged residues D113 and E117 and the anionic pathway is close to R42, R82, and R132. The relevance of the charged residues at the constriction zone is further demonstrated by combined studies on the effect of mutations on these residues. A detailed study including MD simulations and conductance and reversal potential measurements  has demonstrated that mutations generated in the constriction zone by removing cationic residues (R42, R82, and R132) enhance the small cationic selectivity of the channel, whereas removing anionic residues reverses the selectivity \cite{Ulrich2009}. Interestingly, reduction of the negative charge of the constriction zone strongly decreases the conductance of the channel \cite{Ulrich2009}. These effects on the conductivity are observed not only in monovalent electrolyte (KCl)\cite{Ulrich2009} but also in the case of divalent electrolytes (CaCl$_2$) \cite{Miedema2006}. 

Overall, all these results point to the decisive importance of electrostatics at the constriction zone in determining the transport and selectivity properties of OmpF. However, this view has been challenged by the interpretation done of recent reversal potential measurements \cite{Alcaraz2009,Alcaraz2010}. In these experiments, the authors considered the OmpF channel and two of its mutants in several concentrations of different 1:1, 2:1 and 3:1 electrolytes. The two mutants (denoted by OmpF-CC and OmpF-RR) were designed to study the effect of this transversal field by replacing the negatively charged residues located in the constriction zone (D113 and E117) by neutral or positively charged residues. In the CC mutant, both residues were mutated to neutral cysteine residues (D113C and E117C) and in the case of the RR mutant, both residues were mutated into positively charged arginines (D113R and E117R). The resulting constriction zone in the mutant channels is depicted schematically in Figure \ref{channels}b. Reversal potential measurements show a different behavior of the selectivity for each of the channels considered as a function of electrolyte type and concentration. Overall, the results were interpreted by invoking the existence of specific interactions and other (unidentified) structural factors. Our aim here is to perform all-atomic MD simulations of these three ionic channels in order to obtain an atomistic view of the factors involved in the interaction between ions and these channels and its influence on the ionic transport. We will discuss in detail results characterizing the distribution of ions (partition of ions, concentration profiles, 3D isodensity plots) and the transport of ions (cationic and anionic currents) and water (electroosmotic flow) for the OmpF-WT, CC and RR channels.

\section*{Model and Simulation Methods}

\subsection*{Description of the employed Protein Structure}
\label{structure}
Atomic coordinates of the OmpF trimer ionic channel are freely available at the Protein Data Bank database\cite{PDB} (code 2OMF). The downloaded PDB data file contains also spurious atomic coordinates corresponding to atoms from chemical compounds used in the generation of the crystal structures (see details in Ref. \cite{PDB}). Since these coordinates were clearly identified in the PDB file, they were removed before further processing of the PDB file. Also, a few atomic coordinates of the protein structure were missing in the downloaded structure (oxygen atoms for the residue PHE340 and all hydrogen atoms). These missing coordinates from the protein structure were reconstructed using the psfgen structure building module supplied with VMD (version 1.8.6) and NAMD2 (version 2.6) packages \cite{NAMD,VMD}. The resulting trimeric structure is shown in Figure \ref{channels}a, which clearly shows the complexity of the structure of the channel pores. As a definition for the interior of the pore we consider a region 4 nm long given by the $\beta$ barrel structure (3.2 nm long) plus a wider ``entrance'' region (0.4 nm long) situated at each ending of the monomer (note that this definition is close to that employed in previous works\cite{Roux}). The cross section of the channel changes strongly in this region, as discussed in detail already\cite{Roux}. The input structure used in the preparation of the CC and RR mutants was a snapshot of the WT OmpF MD simulation. First all the required residues (Aspartic 113 and Glutamic 117) were mutated using the Mutator module in VMD. In the CC mutant, both residues were mutated to neutral cysteine residues (D113C and E117C) and in the case of the RR mutant, both residues were mutated into positively charged arginines (D113R and E117R).

The protonation states of the residues in the channel employed in the OmpF-WT simulations were determined according to previous detailed studies on the protonation states of OmpF at $p$H=7 \cite{Varma}. In the case of the mutants, changes in the structure can in principle change the pKa of the residues, especially those located near the mutated ones (i.e. in the constriction). We have therefore consistently re-estimated the pKa of all residues using two different methods. First we calculated the pKa of the CC and RR mutants with the program propKa \cite{propKa} and found that, at neutral pH as used in all our simulations, there is no appreciable effect on the protonation states of residues of the protein compared with those in WT OmpF (obviously, with the exception of the mutated residues). We have also calculated the pKa of the mutants using the method described in \cite{Aguilella2004} obtaining the same result. As a consequence, the protonation states of all residues of the mutant OmpF correspond to those used in WT OmpF, except for the mutated residues, whose protonation state corresponds to the pKa obtained from the model pKa, i.e. the standard protonation state at neutral pH.

 The resulting proteins were inserted into a POPC membrane (built with the Membrane Builder supplied with the VMD package \cite{VMD}), removing all lipids in contact with the protein. Water and ions at the desired concentration (and counterions neutralizing the charge of the protein) were inserted using the modules supplied with VMD. 
 The final system, containing the protein embedded in a lipid membrane in a solvated box with electrolyte ions at the desired concentration, was used as input to the whole procedure followed in the MD simulation, including the initial system minimization to remove possible unwanted contacts.

\subsection*{Simulation algorithms and force field}

We employed the software package NAMD version 2.6 for Linux-ia64 \cite{NAMD} to carry out all the Molecular Dynamics simulations reported in the present article. The force-field for protein-lipid simulations supplied by NAMD2 was used, which is a combination of the CHARMM22 force field parameterized to describe protein systems and the CHARMM27 force field, parameterized to describe lipid systems in aqueous media. Water was modeled using the TIP3P model which is the standard choice in the CHARMM force field. 

In all simulations we employed periodic boundary conditions in all directions. Lennard-Jones interactions were computed using a smooth ($10 - 12$ \AA) cutoff, as is standard in NAMD2 simulations. The electrostatic interactions were evaluated using the particle-mesh Ewald (PME) method with a precision of $10^{-6}$ using a $128 \times 128 \times 128$ grid and a $12$ \AA\ cutoff for the real space calculation. This unusually large grid (as compared with previous MD simulations of protein channels, see e.g. refs. \cite{Aksimentiev,Ulrich2009}) was chosen in order to minimize numerical errors leading to spurious effects in spite of its high cost (in terms of computational time). For example, in previous works it was observed significant spurious drift in the center of mass of the system when using less dense grids, an effect which has to be accounted for when computing physical quantities of interest (e.g. ionic currents)\cite{Aksimentiev}. In all our choices we try to ensure very high standards in our calculation of electrostatic interactions which are particularly important in the problem under study.

As described in the next subsections, we have performed simulations under different conditions (NVT, NpT and NVT with external electric field ensembles). In all cases, the  temperature T was controlled by using the Langevin thermostat implemented in NAMD \cite{NAMD}, which is highly efficient for simulations of large systems. The parameters for the thermostat were standard for protein channel simulations; we have employed a damping coefficient of $1$ ps$^{-1}$ and the Langevin forces were applied to all atoms except for hydrogens \cite{Aksimentiev,Ulrich2009}. In the case of NpT simulations, the NAMD2 software controls the pressure by using a Langevin piston Nose-Hoover method (see \cite{NAMD} for details). The value of the pressure applied by piston was set to $p = 1$atm, with a period of the oscillations of $0.1$ps and a relaxation constant of $0.05$ ps. 

In all cases, the equations of motion were solved using a multiple time step in order to speed our very slow simulations, as it is customary with NAMD2 \cite{NAMD}. The values employed for the time step were those usual for the kind of simulations performed here. These are a basic time step of $2$fs and a time step of $4$fs for the evaluation of k-space contribution to the long range electrostatic forces in the PME technique.

We have carefully checked that our choice of parameters correctly reproduce the bulk properties of the electrolytes employed in the simulations (see Ref. \cite{molsim}). In particular, we have found an excellent agreement between simulated and measured values of electrical conductivity and transport numbers for solutions of 1M of KCl. In order  to achieve a good agreement between simulated and experimental values the use of a large simulation box is required, as considered in this work. 

\subsection*{System setup}

The building of the initial configurations for the different simulations was made with the Visual Molecular Dynamics (VMD) Software\cite{VMD}. The OmpF trimer, with its symmetry axis aligned along the $z$ axis, was embedded into a large POPC lipid membrane ($17.71$ nm long both in the $x$ and $y$ axis). In the process, all molecules from the membrane which overlapped with the channel were removed from the system. After that, the system channel-membrane was solvated in a box of preequilibrated TIP3P water molecules and a number of cations (K$^+$) and anions (Cl$^-$) were added to achieve the desired electrolyte concentration. We have constructed three different initial configurations of the channel-membrane system, each one corresponding to a different channel (WT, CC and RR mutants) and ionic solution 1M of KCl. The number of ions and water molecules employed in each case are shown in Table \ref{Table:Sims}. The dimensions of the resulting system were $177.1 \times 177.1 \times 145 $\AA$^3$ for all cases before the equilibration procedure.

\begin{table*}
\caption{\textbf{Size and content of the Simulation box for each simulation}. In this Table we show the length of the simulation box (in nm) in each direction after the NpT equilibration procedure and the number of water molecules, number of ions and total number of atoms employed in each simulation. }
\label{Table:Sims}
\centering
    \begin{tabular}{|l|ccc|c|cc|c|c|}
\hline 
 & \multicolumn{3}{c|}{Size of Simulation Box} & Water & \multicolumn{2}{c|}{Ions} & Atoms & Production \\
  & L$_x$ (nm) & L$_y$ (nm) & L$_z$ (nm) & molecules & K$^+$ & Cl$^-$ &  (Total) & run \\
\hline
OmpF-WT & 17.22 & 17.21 & 14.53 & 108270 & 2033 & 2000 & 446301 & 24.9 ns \\
OmpF-CC & 17.36 & 17.44 & 14.21 & 108264 & 2033 & 2006 & 446274 & 22.93 ns \\
OmpF-RR & 17.36 & 17.45 & 14.21 & 108258 & 2033 & 2012 & 446340 & 23.15 ns \\
\hline
    \end{tabular} 
\end{table*}

\subsection*{Equilibration}

We have employed a carefully designed equilibration procedure, based on previous simulations of ion channels \cite{Aksimentiev,Ulrich2009}. In all simulations, the equilibration procedure has the following steps:

\begin{itemize}
 \item \emph{Energy minimization:} In order to avoid undesired overlaps between atoms, we perform an energy minimization with NAMD of the initial configuration until a constant value for the energy is obtained (typically 2000 steps).
\item \emph{Thermalization of the initial configuration}. The configuration resulting from energy minimization is thermalized by first running a NVT simulation at $T =100$K during 100ps followed by another NVT run at $T=296$K during 100 ps. 
\item \emph{Pressurization of the system} In order to ensure that the system is at a pressure of $p = 1$ atm, we perform two different simulation runs in the NpT ensemble  ($T=296$K, $p = 1$ atm). In the first NpT run, the system was simulated during 1 ns in the isotropic NpT ensemble ($p = 1$atm, $T=296$K), with the protein restrained during the simulation. After this run, a second NpT run is performed during 3 ns. In this case, the protein is unrestrained and the simulation box is allowed to change in size only in the $z$ direction (perpendicular to the membrane plane). The objective of this run is to adjust the pressure of the electrolyte solution in contact with the membrane, as happens in the experimental situation.
\item \emph{Development of ionic current}. After equilibration of the initial configuration, we have performed a run of several ns until an ionic current is developed across the protein channel. These simulations were performed under NVT conditions with an external field with a value equal to that employed in the production runs (see discussion below). 
\end{itemize}

\subsection*{Production runs}
The production runs were performed in the NVT ensemble under conditions of $T=296$K and an external field of $14.22$ mV/nm magnitude pointing in the direction of the negative $z$ axis (see for example Figures \ref{channels} and \ref{crossions}). This value of the electrostatic field corresponds to a potential difference from the top to the bottom of the simulation box of $\sim 200$ mV. Since the resistivity of the electrolyte solutions is extremely low and the resistance of the protein channel is very high, we expect that almost all the potential drop occurs in the region containing the membrane and the protein. The electrical conductivity of 1M KCl electrolyte as obtained from MD simulations of the model employed here is 11.9 S/m (the experimental value is 11.2 S/m) \cite{molsim}. The water electrolyte slab in the simulation for the protein bathed with 1 M KCl is $\approx 10$ nm long (the size of the system excluding the $\approx 4$nm length of the membrane+protein system) and has a cross section of 17$\times$17 nm$^2$, which gives an electrical resistance of about $R\approx 3\times 10^6$ $\Omega$. Previous MD simulations\cite{Ulrich2009} give a conductance for the OmpF channel between 2-3 nS in 1 M KCl, which corresponds to 3-5$\times 10^8$ $\Omega$. Therefore, the electrical resistance of the electrolyte slab is about 2 orders of magnitude smaller than the electrical resistance of the channel, so it can be safely assumed that the drop of electrostatic potential in the electrolyte solution is negligible as compared with the drop across the protein+membrane system. A discussion on this methodology (application of an external uniform field for calculation of ionic conductivies in protein channels) is discussed in detail in Ref \cite{Rouxmania}.

We note here that previous MD simulations\cite{Aksimentiev,Ulrich2009} of protein channels were performed by using much larger values for the electrostatic potential drop (around 1 V) in order to obtain good statistics for the flow of ions. Since these high potentials cannot be obtained experimentally, we have decided to employ a smaller electric field in order to obtain a potential drop which is realistic yet gives a ion flux which can be observed in long simulations. The duration of the production runs was selected by ensuring that good statistics in the ion flow were obtained (see next section). In all cases, very long simulations were required, as shown in Table 1. It has to be emphasized that, due to the large number of atoms in the simulation box and the high precision of the calculations, our production runs are extremely slow, so the simulations runs lasted during several months. Typically, we required around 1.12 days/ns using 64 processors ``Itanium Montvale'' at the CESGA Supercomputing facility.

\subsection*{Calculation of ionic currents}

In the production runs we have computed the electric current by using three independent methods. Inconsistencies between the results obtained by these three methods allow us to detect more easily possible inacuracies in the simulations such as those due to poor sampling. 
\begin{itemize}
 \item \emph{Method 1}: crossing ions. In this method, we have monitored the number of cations which , at time $t$, completely crossed the protein in the direction of the electric field, $N_{\text{K}^+}^{cr}(t)$. Also, we have computed the number of anions crossing the protein in the opposite direction $N_{\text{Cl}^-}^{cr}(t)$. These numbers of ions were counted by following the individual trajectories of each ion. An ion was considered to cross the channel if it is initially at one side of the membrane and ends at the opposite side of the membrane bulk electrolyte following a path inside the protein channel (some ions were found to enter inside the channel without crossing the constriction region, returning again to their initial side of the membrane, these ions were not counted in the number of ions crossing the channel). The total charge from anions and cations that has flowed accross the protein channel up to time $t$ is given by:
\begin{equation}
\label{method1}
 Q^f(t) = e N_{\text{K}^+}^{cr}(t) - (-e)N_{\text{Cl}^-}^{cr}(t)
 \end{equation}
In the stationary state, this amount of charge is related to the current intensity $I$ by $Q^f(t)=I\times t$. In practice, we compute $Q^f(t)$ during the simulation runs and the intensity is obtained from a linear fit (see raw data and resulting fits in the Online Supporting Material).

\item \emph{Method 2}: total intensity. In this method, one takes into account the current flowing across all the system \cite{Aksimentiev} by noting that in the stationary state, the current flowing in the aqueous electrolyte region has to be equal to the current flowing throw the channel. In this method, the instantaneous current is computed by \cite{Aksimentiev}:
\begin{equation}
\label{Ieq}
 I(t) = \frac{1}{\Delta t L}\sum_{i =1}^{N}q_i\left[ z_i(t+\Delta t) - z_i(t)\right],
\end{equation}
where $z_i$ and $q_i$ are the $z$-coordinate and the charge of atom $i$, respectively. $L$ is the size of the simulation box and $\Delta t$ is the time interval employed to record data, which was chosen to be $\Delta t = 10$ps. The average current is computed by linearly fitting the cumulative current that is obtained by integration of the instantaneous current given by Eq.(\ref{Ieq}). 

\item \emph{Method 3}: intensity across a fixed plane. A simple method to obtain the current intensity is based on counting the number of cations ($N_{\text{K}^+}^{(p)}$) and  anions ($N_{\text{Cl}^-}^{(p)}$) crossing a plane perpendicular to the direction of the electric field. In the stationary state, this current has to be independent of the selected plane. In our calculations, we have selected the plane $z=0.5$ nm which corresponds to the narrower part of the constriction zone. The total charge from anions and cations that has flowed accross the selected plane up to time $t$ is given by:
\begin{equation}
\label{method3}
 Q^{p}(t) = e N_{\text{K}^+}^{(p)}(t) - (-e)N_{\text{Cl}^-}^{(p)}(t)
 \end{equation}
In the stationary state, this amount of charge is related to the current intensity $I$ by $Q^p(t)=I\times t$. In practice, we compute $Q^p(t)$ during the simulation runs and the intensity is obtained from a linear fit (see raw data and resulting fits in the supporting Online Auxiliary Material).

\end{itemize}
\section*{Results and Discussion}

\subsection*{Partition of ions between channel and electrolyte}
In Table \ref{Table:Nions}, we show the results for the number of ions inside the channel. Our results show that the ionic charge inside the channel (the charge from Cl$^-$ and K$^+$) is different for each protein. In fact, the differences in ionic charge inside the mutants and the wild type channel can be explained by the change in protein charge induced by the mutations. The mutations of the CC and RR channels increase the charge of each protein monomer (as compared with the WT case) by an amount of +2$e$ and +4$e$ respectively. The results in Table \ref{Table:Nions} nicely illustrate the electrostatic balance inside the channel: the changes in the protein charge due to mutations induce (within the uncertainty of the simulations) changes of the same magnitude and opposite sign in the ionic charge inside the protein channel. This could be interpreted as a test for the validity of overall electroneutrality in the channel-solution system as assumed in simplified treatments based on effective charge \cite{Aguilella2007}.

It is also remarkable that in all three channels there is an excess of cations over anions (see Table \ref{Table:Nions}). This result can be understood form the fact that the overall charge of all channels is negative. In our simulations of OmpF-WT at 1 M KCl, we observe an approximate ratio of 1.7 K$^+$ for each Cl$^-$ inside the channel. In the case of the OmpF-CC mutant, the protein still has a clear preference to be occupied by cations over anions; we obtain a ratio of 1.42 K$^+$ for each Cl$^-$. In the case of OmpF-RR, there is only a slight preference over cations, we obtain a ratio of only 1.13 K$^+$ for each Cl$^-$.

\begin{table}
\caption{Number of ions of each species inside the trimer protein channel averaged over the simulation runs (statistical errors in the mean are estimated from $2\sigma$). From this data, we also compute the ionic charge $\Delta Q_\text{ions}$ inside each channel compared with the OmpF-WT case. For comparison, we also give the charge of each protein channel $\Delta Q_\text{channel}$ with respect to the OmpF-WT case. All simulations correspond to 1M KCl.}
\label{Table:Nions}
\centering
\begin{tabular}{|l|cc|cc|}
\hline
           & \multicolumn{2}{c|}{Channel Occupancy} & \multicolumn{2}{c|}{Charge respect to WT} \\
        & Cl$^-$ & K$^+$ & $\Delta Q_\text{ions}$ & $\Delta Q_\text{channel}$ \\ \hline
     OmpF-WT & $21.0 \pm 0.2$ & $35.5 \pm 0.2$ & - & - \\
     OmpF-CC & $23.2 \pm 0.1$ & $33.1 \pm 1.4$ & $(-5\pm 2)e$ & $+6e$ \\
     OmpF-RR & $30.4 \pm 0.4$ & $34.4 \pm 1.4$ & $(-11\pm 2)e$ & $+12e$ \\
\hline  
    \end{tabular} 
\end{table}

\begin{figure}[htp]
       \includegraphics*[width=14cm]{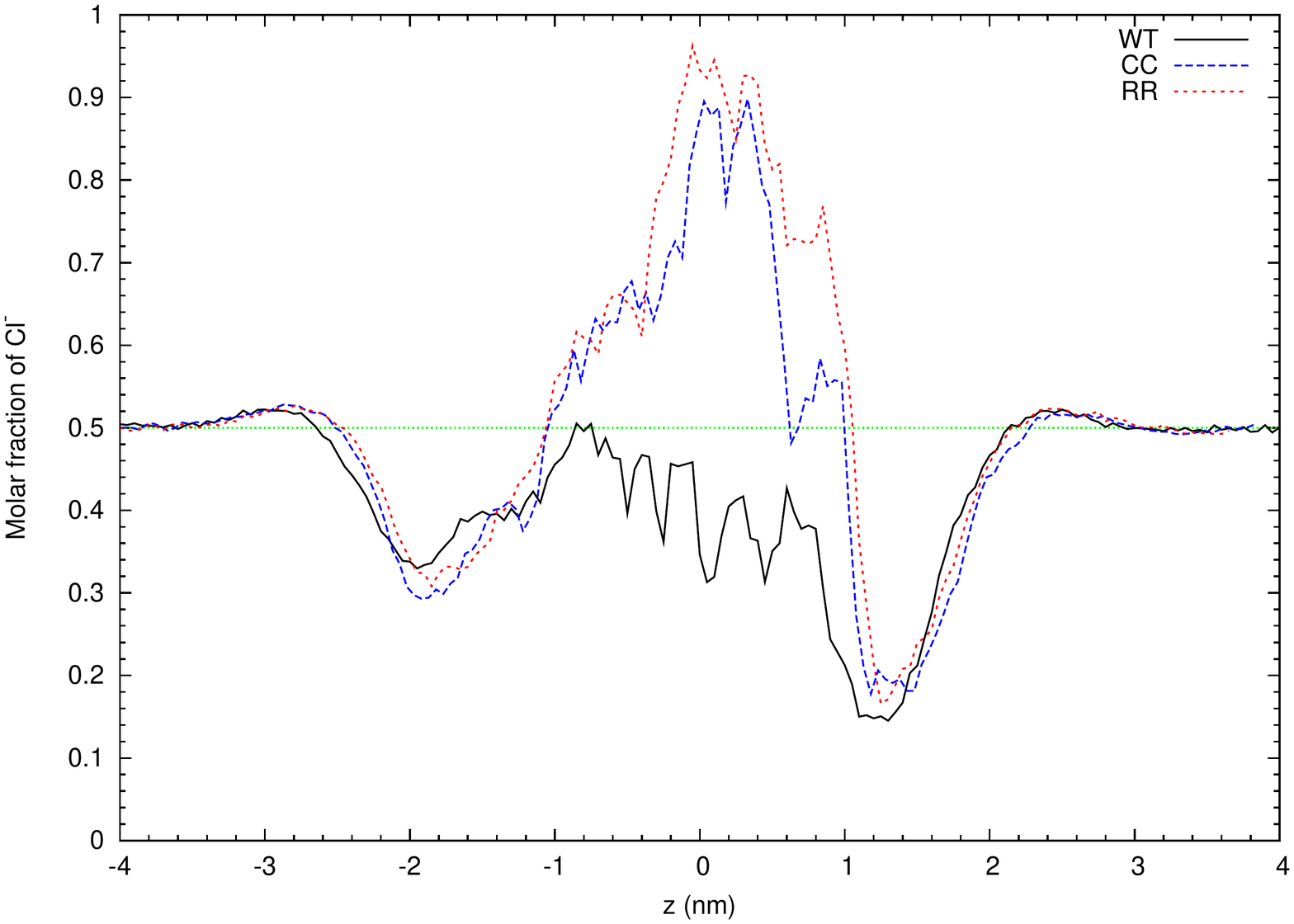}
       \caption{\label{profile} Average profile of the fraction of anions (Cl$^-$) with respect total number of ions (Cl$^-$ and K$^+$) as a function of the axial coordinate $z$ (we show only the central region of the simulation box).}
 \end{figure}

At this point, it is important to remark that the distribution or partition of the ions inside the different regions of the protein channel shows striking differences, as should be expected from structural arguments. This can be clearly appreciated in Figure \ref{profile}. In the case of the OmpF-WT channel, there is a clear excess of K$^+$ over Cl$^-$ in all the regions of the channel. On the other hand, the two mutants show a clear exclusion of K$^+$ near and at the constriction region. This can be clearly observed in the peaks shown in figure \ref{profile} which correspond to the constriction region. At this peak, the ratio of anions over the total number of ions is 0.90 in the case of the CC mutant and 0.96 in the case of the RR mutant. These striking differences between OmpF-WT and the mutants can be attributed to the fact that the WT channel has both positively charged and negatively charged residues at the constriction zone, whereas the mutants  do not have negatively charged residues at this region (see Figure \ref{channels}).  

\begin{figure}[htp]
   \begin{center}   
       \includegraphics*[width=12cm]{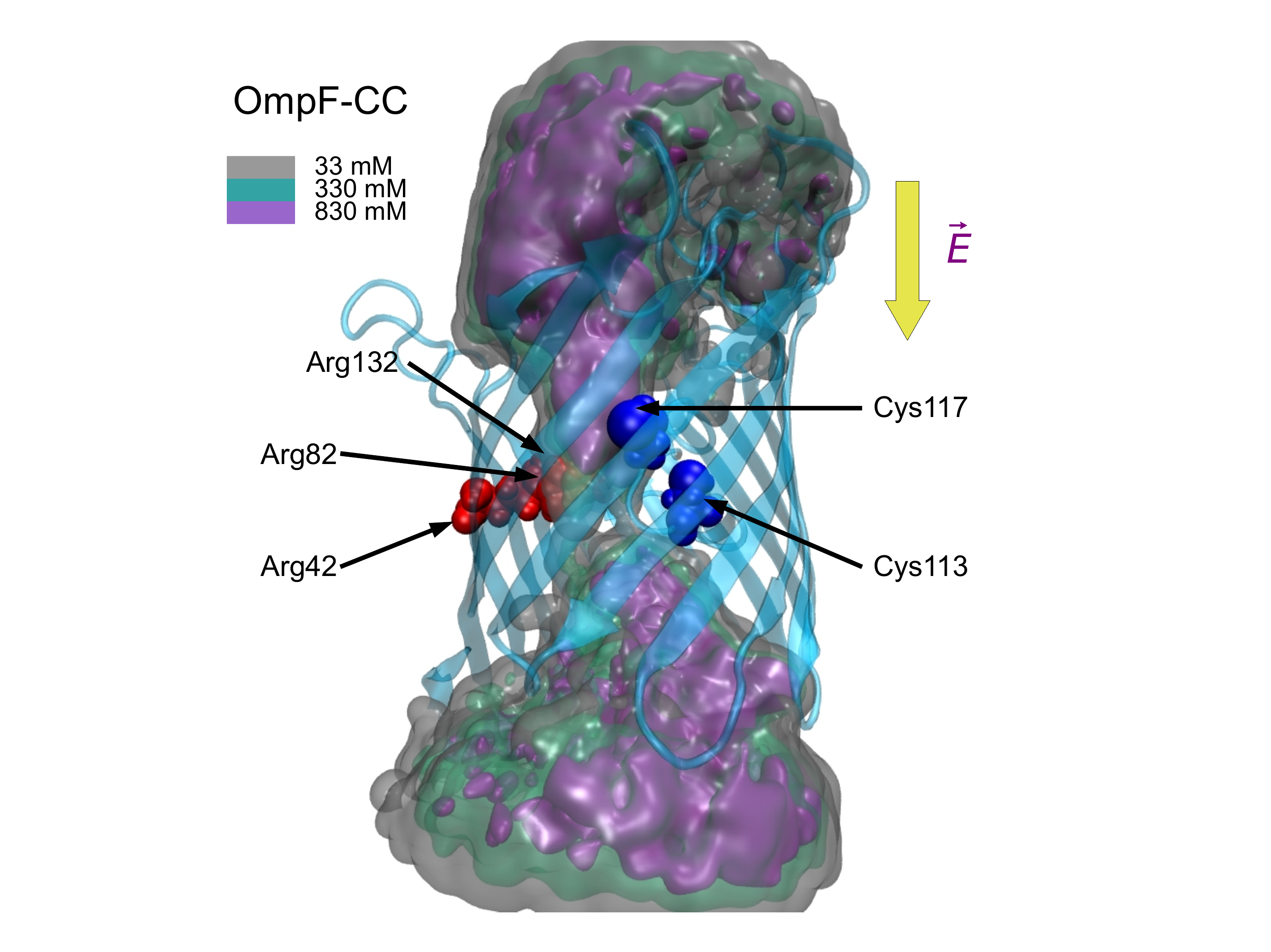}
       \includegraphics*[width=12cm]{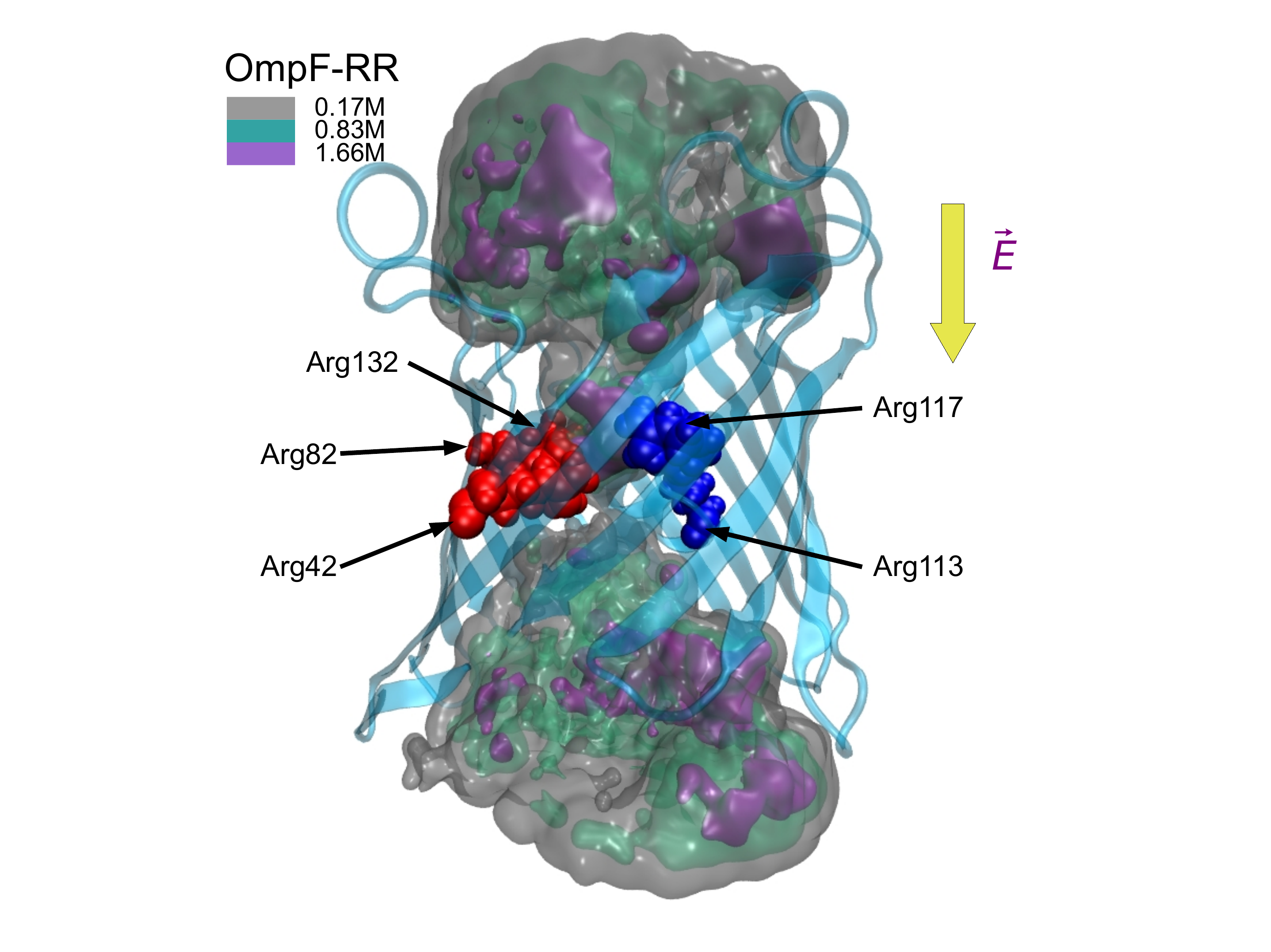}
       \caption{\label{isodensityK} Iso-density surfaces for the K$^+$ ion inside the CC and RR mutants. Three different isodensity values are shown in order to demonstrate the low K$^+$ concentration observed past the region delimited by the mutated residues. The figures were generated using VMD \cite{VMD}.}
    \end{center}
 \end{figure}

\begin{figure}[htp]
   \begin{center}   
       \includegraphics*[width=12cm]{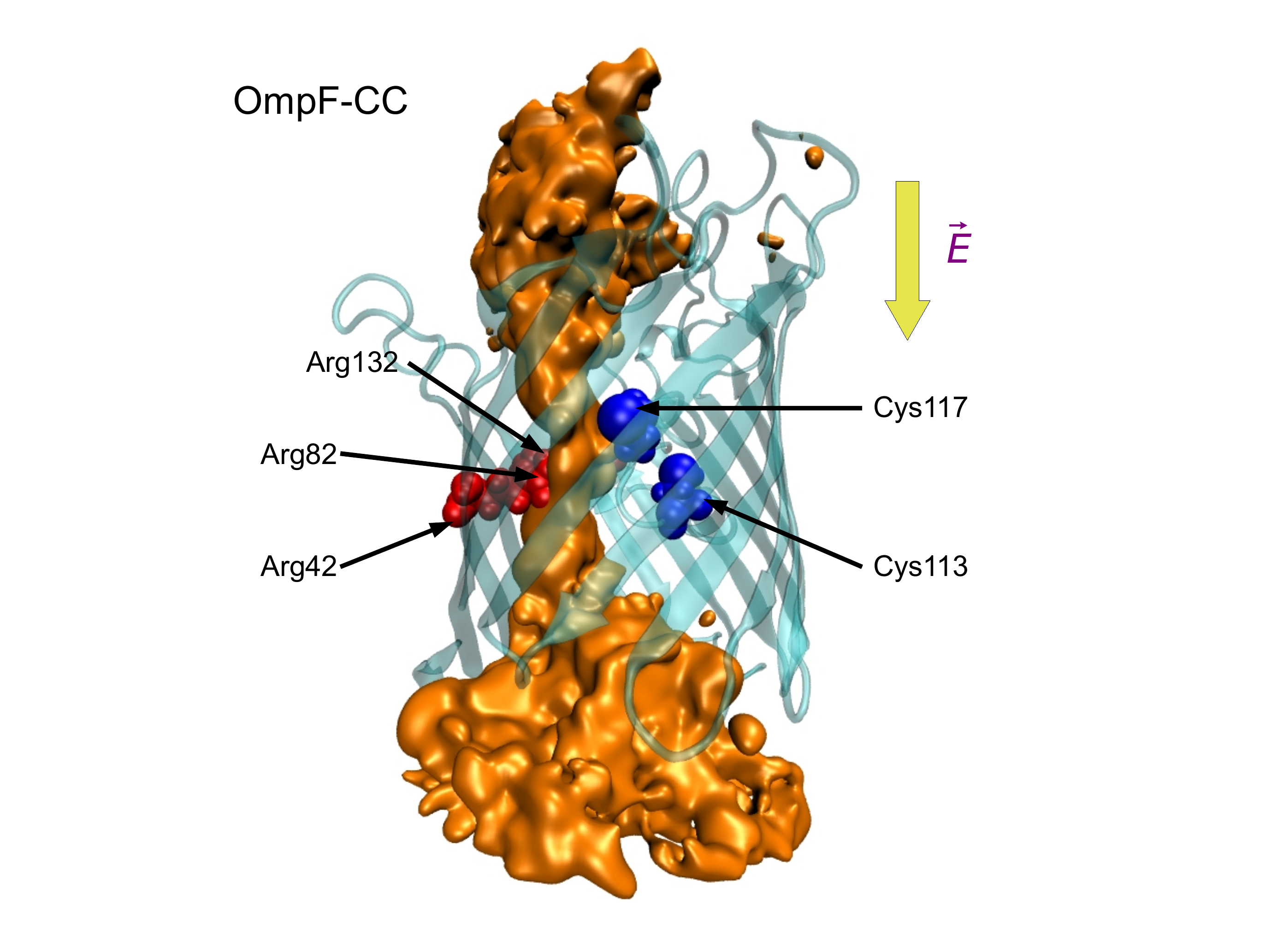}
       \includegraphics*[width=12cm]{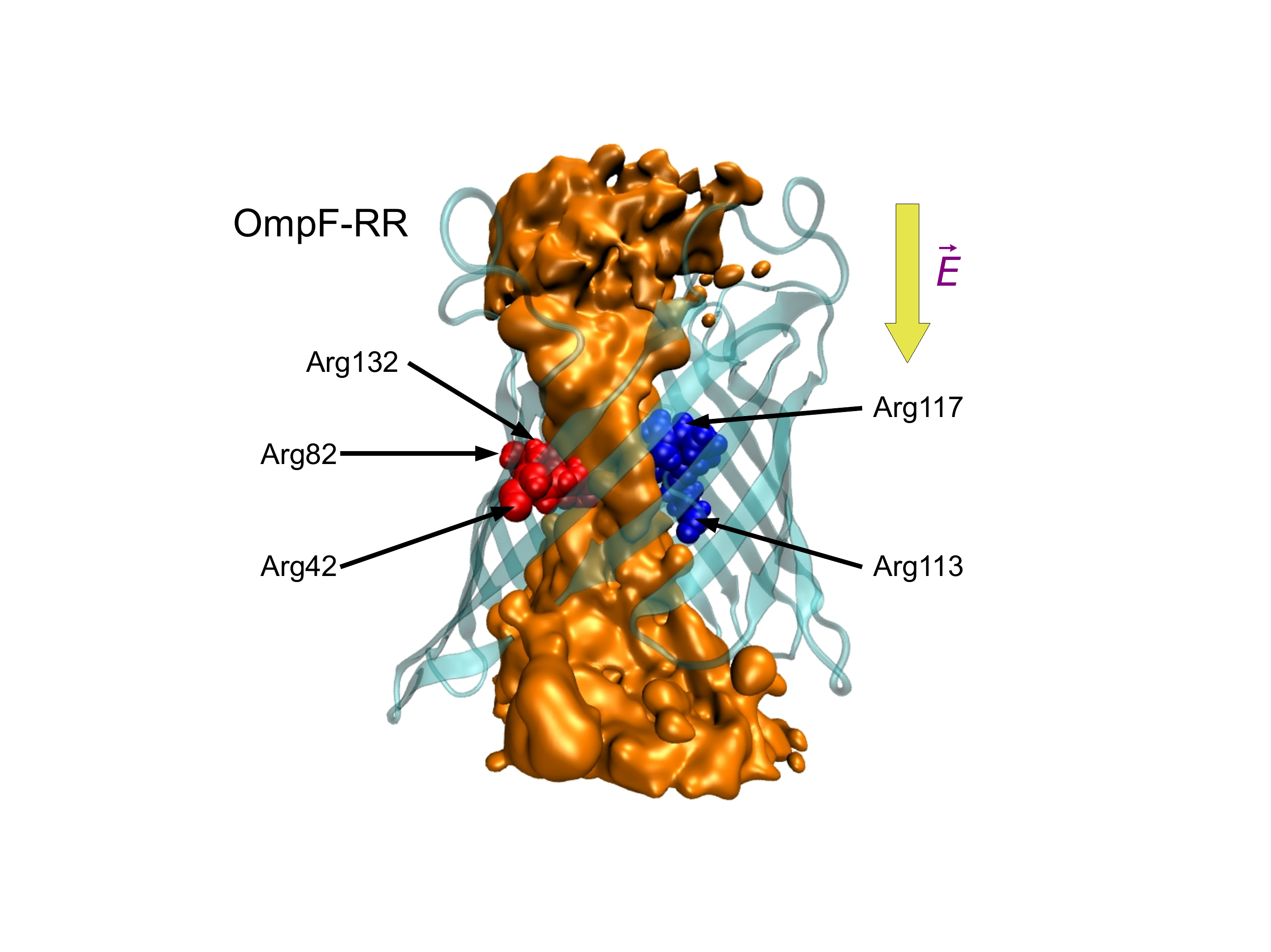}
       \caption{\label{isodensityCl} Iso-density surface corresponding to 0.6 M concentration of the Cl$^-$ ion inside the CC and RR mutants. The figures were generated using VMD \cite{VMD}.}
    \end{center}
 \end{figure}

In fact, in the case of the OmpF-WT channel, the positively charged residues R42, R82 and R132 and the acidic residues D113 and E117 create a transversal field which causes anions and cations to cross the constriction region by different pathways. The existence of these pathways was predicted from different simulation techniques \cite{Roux} and it has been recently confirmed by anomalous X-ray scattering \cite{Roux2010}. Since the transversal field of the OmpF-WT constriction is greatly perturbed in the CC and RR mutants (in which the acidic residues are replaced by neutral or positively charged residues, see Figure \ref{channels}) we should not expect the existence of these pathways. This can be appreciated in Figures \ref{isodensityK} and \ref{isodensityCl}. First, note that the concentration of cations is very low past the constriction zone. Therefore, the cationic selective pathway present in OmpF-WT has been suppressed by the CC and RR mutations which remove the negative charges from the constriction zone. Concerning the distribution of anions inside the channel (Figure \ref{isodensityCl}), the effects of the mutations are also significant. In the case of the CC mutant, anions  still follow a definite path close to the positively charged residues R42, R82 and R132 and are not significantly found near C113 and C117 which are neutral. Other studies with a mutant similar to CC (neutralization of the negative residues with Asparagine and Glutamine instead of Cysteine, D113N and E117Q) shows also the same behavior \cite{Ulrich2009}.

In the case of the RR mutant, anions are found with the same concentration in all the available space of the constriction region. All residues of the constriction zone are positively charged (R42, R82, R132, R113 and R117) so the anions do not show any preference for particular sides of the walls of the protein channel.

\subsection*{Flow of ions across the channel}
\begin{figure}[htp]
   \begin{center}   
       \includegraphics*[width=10cm]{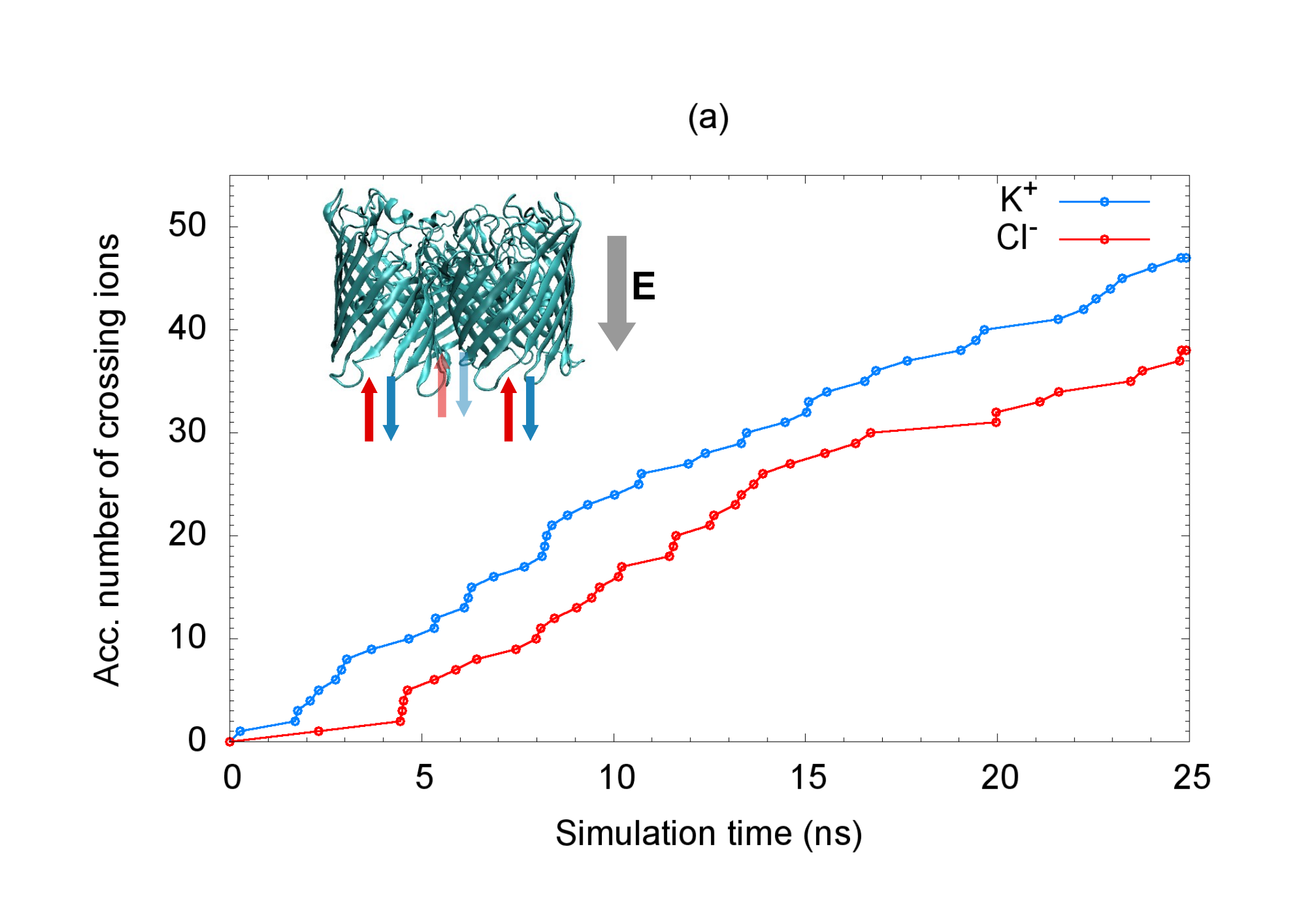}
       \includegraphics*[width=10cm]{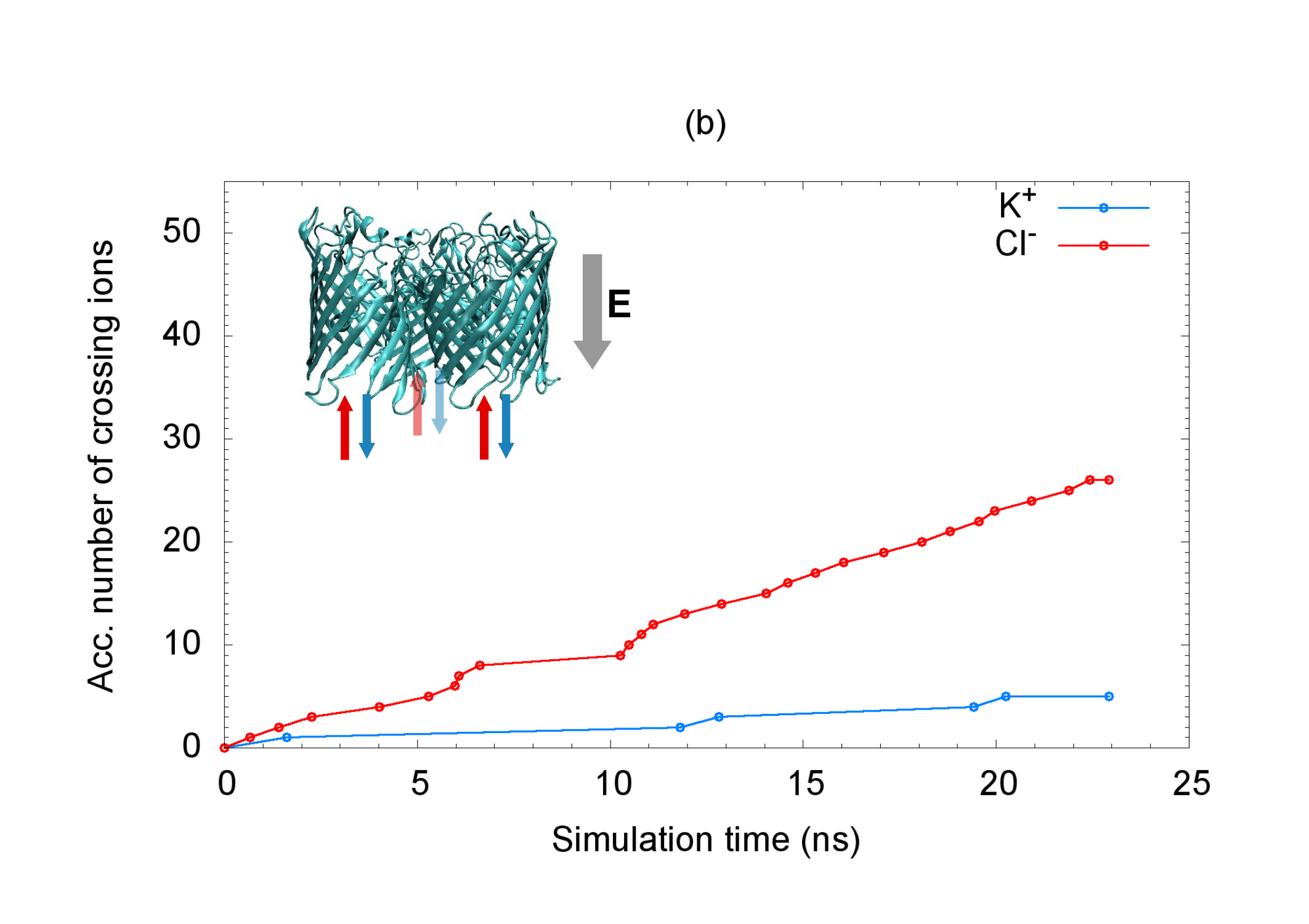}
       \includegraphics*[width=10cm]{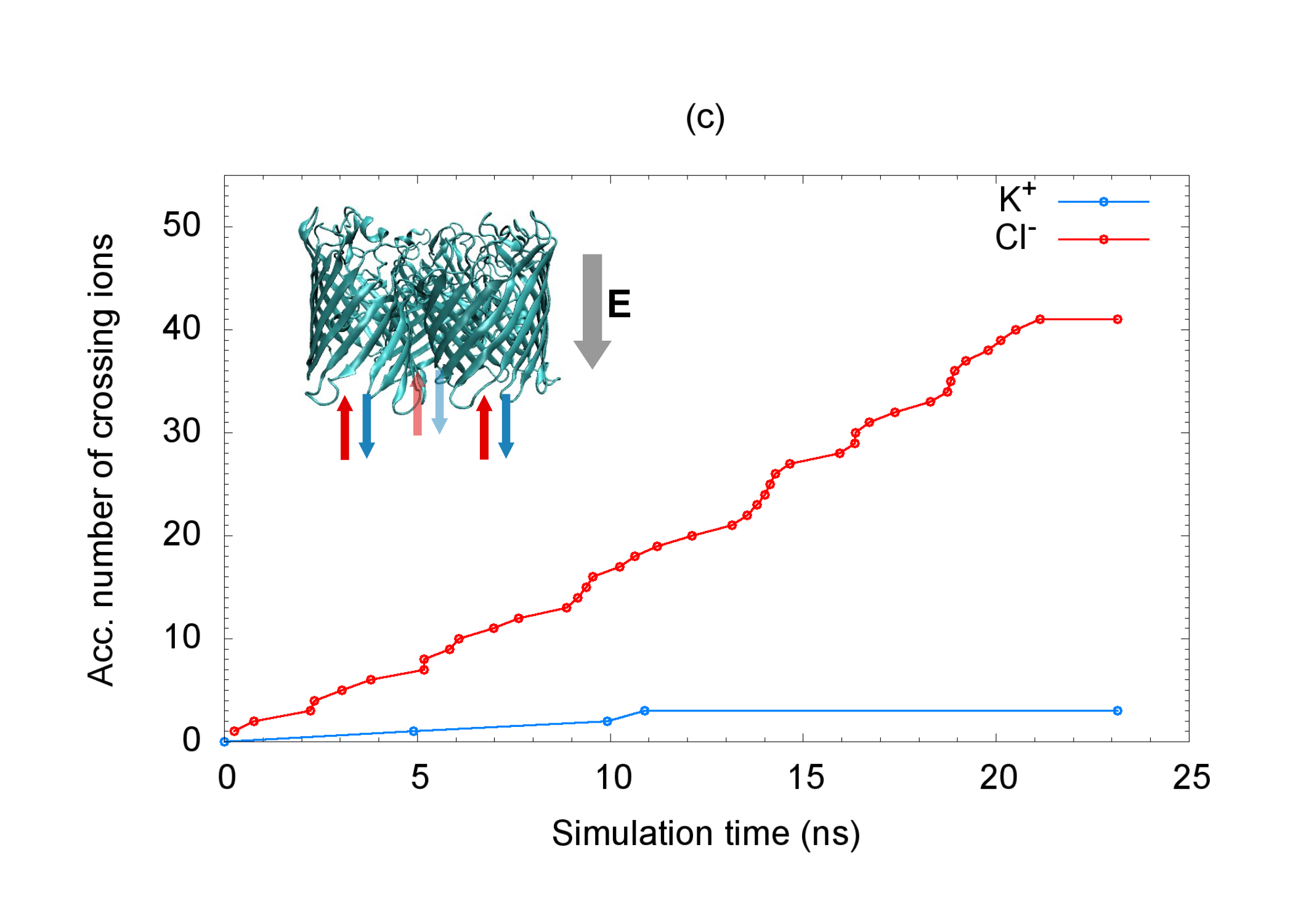}
       \caption{\label{crossions} Accumulated number of ions crossing the protein channel as a function of time (a) OmpF-WT (b) OmpF-CC (C) OmpF-RR }
    \end{center}
 \end{figure}

In order to study the effect of mutations in the transport of ions, we have computed the number of ions with a trajectory that crosses the protein channel, as described in the Methods section (see also previous MD studies of the OmpF channel \cite{Ulrich2009}). The results are shown in Figure \ref{crossions}. In agreement with previous results, the number of cations crossing the OmpF-WT channel is slightly larger than the number of anions. We observe a ratio of 1.24 between the events of K$^+$ and Cl$^-$ ions crossing the channel, consistent with previous MD results \cite{Ulrich2009}.

In the case of the two mutants (CC and RR), the large majority of ions crossing the channel are anions. In fact, only a few events of cations crossing the mutant channels were observed (5 in the CC case and 3 in the RR case). It is also interesting to note that the amount of anions crossing the channel per unit time is similar in the WT and RR case but it is significantly smaller in the CC case. In fact in the case of the CC and RR mutants we see almost only anions crossing the channel, despite the fact that there are significant amounts of cations inside the channel (see Table 2). However, as shown in Figures \ref{profile},\ref{isodensityK} and \ref{isodensityCl}, cations are expelled from the positively charged constriction zone (at which we find almost only anions). In these mutants, there is a large  electrostatic barrier at the constriction, making extremely difficult the crossing of cations across the channel.

The previous discussion suggest a strong reduction of the current due to the effect of the mutations. In order to evaluate this effect, we have calculated the current intensity using the three different methods described in the previous section (see Supplementary Online Material for details). Let us remark that the results obtained from these different methodologies (Table \ref{Intensity}) show an excellent consistency, substantially improved from previous works using smaller system sizes and lower resolution in the PME electrostatic calculation. The equivalence between the three methods for the calculation of the current intensity indicates that the sampling and precision of our calculations were sufficiently accurate for the calculation of ionic currents altough at the expense of highly demanding simulations.

Our MD results predict a substantial reduction of the current intensity for the CC and RR mutants as compared with the WT channel. The results are consistent with a reduction of the conductance for the CC mutant in 1 M KCl in a factor $\sim 2.5$ as compared with the case of OmpF-WT. According to our previous discussion, this reduction can be interpreted as due to the blocking of the flow of cations and a small reduction in the flux of anions (see Figure \ref{crossions}). Interestingly, conductance measurements in different conditions also show a reduced conductance of OmpF-CC with respect to OmpF-WT. At conditions of pH = 5.7,� $T$ = 296 K and 2 M concentration of KCl, measurements give for the conductances 7.18 nS for OmpF-WT and 2.12 nS for OmpF-CC \cite{Review}. A  similar reduction in conductance is also observed for different conditions in CaCl$_2$ \cite{Miedema2006}. Other studies with similar mutants give also the same tendency. For example, in the case of neutralization of the negative residues with Asparagine and Glutamine instead of Cysteine (D113N and E117Q) \cite{Ulrich2009} both experiments and MD simulations predict a strong reduction of the conductance (by a factor $\sim$2 in presence of 1 M KCl). 

In the case of the RR mutant, the intensity of the current is reduced in a factor $\sim 1.5$ as compared with the WT case. In this case, we also observe a blocking of the flow of cations (see Figure \ref{crossions}c) but we also observe an increase in the flow of anions as compared with the CC case. From the structural point of view this is quite reasonable, since the mutations of the RR channel facilitate the translocation of anions across the constriction zone (recall for example the iso-density plots discussed previously, Figure \ref{isodensityCl}). Experimental results are also consistent with a smaller reduction in conductance from WT to RR than that observed in the CC case \cite{Miedema2006}.

\begin{table}
\caption{Current intensity flowing in the different protein channels under a potential drop of 200 mV bathed by 1 M KCl at $T$=296K computed from MD simulations using three different methods (see Methods section). }
\label{Intensity}
    \begin{tabular}{|l|ccc|}
\hline 
 & \multicolumn{3}{c|}{Current intensity} \\

        & Method 1 & Method 2 & Method 3 \\ \hline
OmpF-WT & 0.57 nA & 0.53 nA & 0.53 nA \\
OmpF-CC & 0.21 nA & 0.20 nA & 0.20 nA \\
OmpF-RR & 0.32 nA & 0.35 nA & 0.33 nA \\
\hline
    \end{tabular} 
\end{table}

\subsection*{Electroosmotic flow of water}
\begin{figure}[htp]
   \begin{center}   
       \includegraphics[width=9.5cm]{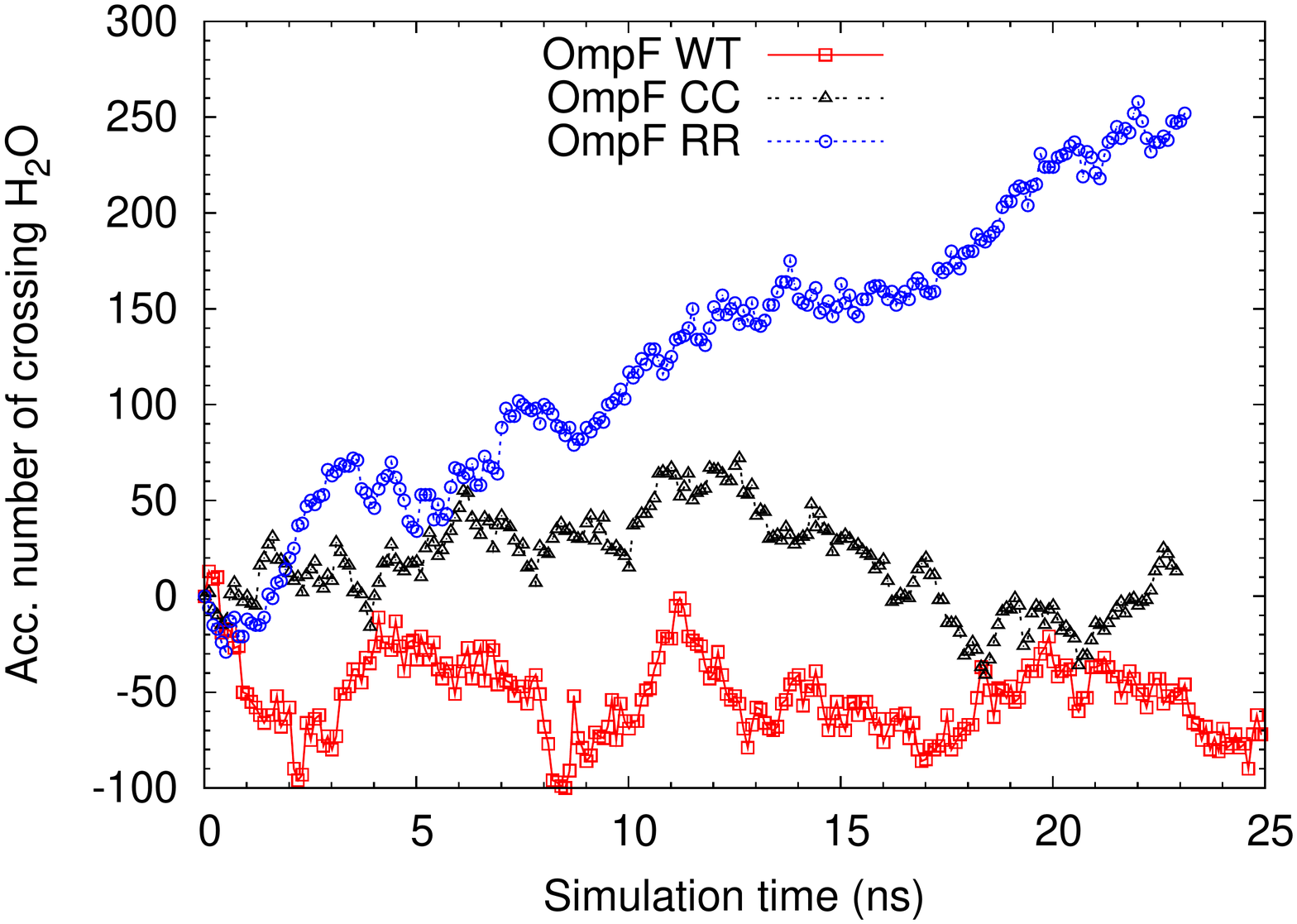}
       \caption{\label{water} Accumulated number of water molecules crossing the OmpF-WT channel (squares), and the two mutants OmpF-CC (triangles) and OmpF-RR (circles). A negative value indicates a flow in the direction of negative $z$ (which is also the direction of the applied electric field). }
    \end{center}
 \end{figure}

Previous MD simulations of other protein channels \cite{Aksimentiev} have shown the existence of a flow of water accompanying the transport of ions. We have also analyzed this possibility for the three different protein channel considered here. There is only one case with a significant flow of water, which corresponds to the OmpF-RR mutant (see Fig. \ref{water}). In the case of the RR mutant, we observe a substantial flow of $\sim$ 10-11 water molecules/ns crossing the protein channel in a direction opposite to that of the applied external field. This water flow has a direction and magnitude consistent with a transport of hydration water by the anions crossing the channel. In our simulation, we observe a ratio of 6.25 water molecules crossing the channel per each anion also crossing the OmpF-RR channel. This ratio is only slightly smaller than the average hydration number of Cl$^-$ in 1 M KCl, which is 7 water molecules per Cl$^-$ in our model of electrolyte \cite{molsim}.

As we said previously, the flow of water is negligible in the case of the OmpF-WT protein and the CC mutant. Our results suggest that only in the case of the RR mutant the field present in the constriction zone allows charge carriers (anions in this case) to cross the channel without perturbing its hydration shell.

\section*{Conclusions}
Our present study gives further support to the concept that MD Simulation studies are useful to understand the effect of protein structural factors (and therefore the effect of mutations) in the transport of ions across nanochannels. Our results also point out to the necessity of combining analysis of transport properties (ionic fluxes) and the details of partition of ions between the ionic solution and the protein interior. Such quantities provide different pieces of information related in a different way to the structural properties of the mutated channels. In the cases studied here, average numbers for the partition of ions between the electrolyte solution and the protein channels simply reflect the effective charge of the channels. The details of the distribution of ions inside the protein are rather complex, which is caused by the local structure of the nanochannel (for example, excluding cations from a positively charged constriction zone). In addition, ionic fluxes are clearly determined by the charge of the constriction zone and do not seem to be correlated with partition numbers.

MD simulations allow the study of the behaviour of water molecules inside the protein channel, in contrast with less detailed approaches. In the case of OmpF-WT and OmpF-CC mutant we have found that the flow of water is negligible. However in the case of the OmpF-RR mutant this is not longer true. In that case we obtain a substantial flow of water consistent with the hydration shell of the crossing Cl$^-$.

The great advantage of the MD simulation technique is the wealth of structural and transport information provided with atomistic detail. However, this technique has a severe limitation in its high computational cost, which is clearly seen in our simulations discussed here. In order to obtain realiable results for the ionic currents, we needed to perform long runs in large systems. This limitation forced us to study only a very limited set of conditions (1M KCl in three different channels), instead of exploring a wide range of concentrations and mutations. These limitations are even more severe in the case of multivalent ions. However, the study of the interaction between multivalent electrolytes and the protein channels is of great interest, as emphasized by experimental results showing a highly nontrivial behaviour \cite{Alcaraz2009,Alcaraz2010}. Work in the direction of understanding the role of multivalent ions and relate it to newly found phenomena in the field of electrokinetics is under way \cite{Aguilella2010}.

\section*{Acknowledgments}
This work is supported by the Spanish Government (grants FIS2009-13370-C02-02, FIS2007-60205 and CONSOLIDER-NANOSELECT-CSD2007-00041), Generalitat de Catalunya (2009SGR164) and Fundaci\'o Caixa Castell\'o-Bancaixa (P1-1A2009-13). C.C. is supported by the JAE doc program of the Spanish National Research Council (CSIC). The Supercomputing resources employed in this work were provided by the CESGA Supercomputing Center, Spain. We acknowledge useful discussions with V.M. Aguilella and A. Alcaraz.


\end{document}